\RequirePackage{fix-cm}

\documentclass[pdftex,twocolumn,epjc3]{svjour3}  
\usepackage{amssymb}
\RequirePackage[T1]{fontenc}
\RequirePackage{mathptmx}      
\RequirePackage{flushend}
\DeclareMathAlphabet{\mathcal}{OMS}{cmsy}{m}{n} 
\smartqed
\pdfoutput=1 
\usepackage{graphicx,epsfig}
\usepackage{caption}
\usepackage{subcaption}
\usepackage{mathrsfs}
\usepackage{dcolumn}
\usepackage{bm}
\usepackage{amsmath}
\usepackage{epstopdf}
\usepackage{amsfonts}
\usepackage{amssymb}
\usepackage{xcolor}
\usepackage{float}
\usepackage{multirow}
\usepackage{appendix}
\usepackage{color}
\usepackage{relsize}
\usepackage{xparse}
\usepackage{accents}

\newcommand{\erf}{\operatorname{erf}}

\newcommand{\be}{\begin{equation}}
\newcommand{\ee}{\end{equation}}
\newcommand{\bea}{\begin{eqnarray}}
\newcommand{\eea}{\end{eqnarray}}

\newcommand{\avg}[1]{\langle #1\rangle}

\journalname{Eur. Phys. J. C}
\begin{document}

\title{Classical and quantum cosmology of $f(R)$ gravity's rainbow in Schutz's formalism
}


\author{Andrés Burton-Villalobos\thanksref{e1,addr1}
        \and Giovanni Otalora \thanksref{e2,addr1}
        \and
      Manuel Gonzalez-Espinoza \thanksref{e3,addr2}
        \and 
       Yoelsy Leyva\thanksref{e4,addr1}
        }


\thankstext{e1}{e-mail: andres.burton.villalobos@alumnos.uta.cl}
\thankstext{e2}{e-mail: giovanni.otalora@academicos.uta.cl}
\thankstext{e3}{e-mail: manuel.gonzalez@pucv.cl}
\thankstext{e4}{e-mail: yoelsy.leyva@academicos.uta.cl}


\institute{Departamento de Física, Facultad de Ciencias, Universidad de Tarapacá, Casilla 7-D, Arica, Chile\label{addr1}   \and  Instituto de F\'{\i}sica, Pontificia Universidad Cat\'olica de
Valpara\'{\i}so, Casilla 4950, Valpara\'{\i}so, Chile.  \label{addr2}    
}

\date{Received: date / Accepted: date}

\maketitle

\begin{abstract}
We investigate the classical and quantum dynamics of $f(R)$ gravity's rainbow in the presence of a perfect fluid, employing Schutz's formalism to establish a well-defined notion of time. In the classical regime, we derive and solve the equations of motion, obtaining both analytical and numerical solutions. Through canonical quantisation, we formulate the Schr\"{o}dinger–Wheeler–DeWitt (SWD) equation for the quantum model. By solving its eigenfunctions, we construct the wave function of the Universe and obtain analytical solutions in scenarios dominated by stiff matter. Our results highlight the impact of rainbow gravity on quantum evolution, particularly in modifying the structure of the wave function and shaping the transition from the quantum to the classical regime.
\end{abstract}

\section{Introduction}\label{Introduction}
Shortly after the establishment of the General Theory of Relativity (GR), scientists began exploring modifications to its framework. In 1919, Weyl \cite{Weyl:1919fi} introduced higher-order curvature invariants into the Einstein-Hilbert action, while Nordstr\"{o}m \cite{Nordstrom:1913dga} proposed a conformally flat scalar theory of gravity, representing some of the earliest alternatives to GR. Later, building on Dirac's 1937 idea of a varying gravitational constant \cite{Dirac:1937ti}, Brans and Dicke formulated a scalar-tensor theory in 1961, incorporating a scalar field alongside the metric tensor to mediate gravitational interactions \cite{Brans:1961sx}. The need for such modifications was further reinforced by the non-renormalizability of GR. In 1962, Utiyama and DeWitt demonstrated that renormalization at the one-loop level requires supplementing the Einstein-Hilbert action with higher-order curvature terms \cite{Utiyama:1962sn}. Furthermore, quantum corrections from string theory have shown that the effective low-energy gravitational action naturally includes such higher-order invariants \cite{Birrell:1982ix,Buchbinder:1992rb,Vilkovisky:1992pb}.  

This strong motivation to extend GR has given rise to modified gravity theories, with $f(R)$ gravity being one of the simplest and most extensively studied proposals \cite{DeFelice:2010aj,Nojiri:2017ncd,Nojiri:2010wj}. A major advantage of $f(R)$ gravity is its ability to avoid the Ostrogradski instability, a fundamental issue in most higher-order theories \cite{Woodard:2006nt}. A prominent example is Starobinsky's $R^2$ inflation model \cite{Starobinsky}, which explains the Universe's rapid accelerated expansion shortly after the Big Bang. This inflationary period occurred approximately $ 10^{-34}$ seconds after the singularity and lasted between $10^{-33}$ and $10^{-32}$ seconds, while also accounting for the generation of primordial fluctuations \cite{Guth:1980zm,Linde:1981mu,Albrecht:1982wi}. Moreover, although initially motivated by high-energy physics near the Planck scale, the discovery of late-time cosmic acceleration in 1998 further renewed interest in $f(R)$ gravity \cite{DeFelice:2010aj,Nojiri:2017ncd,Nojiri:2010wj} and related models such as $f(T)$ gravity \cite{Linder:2010py,Bengochea:2008gz,Cai:2015emx}.  

Another intriguing extension of GR is Gravity's Rainbow, which introduces small-scale ultraviolet (UV) modifications while preserving GR as the low-energy limit \cite{Magueijo:2002xx}. Unlike other quantum gravity frameworks, these UV modifications directly affect the spacetime metric by extending double special relativity to curved spacetimes. In Gravity's Rainbow, the metric tensor depends on the energy of particles probing spacetime \cite{Magueijo:2002xx,Amelino-Camelia:2013wha,Assanioussi:2014xmz,Assanioussi:2016yxx}. A natural extension of this framework is $f(R)$ Gravity's Rainbow, which introduces additional degrees of freedom, enabling richer cosmological dynamics while still incorporating the energy-dependent spacetime deformations characteristic of Gravity's Rainbow. Studies on $f(R)$ Gravity's Rainbow have examined inflationary models, including Starobinsky inflation with rainbow functions linked to the Hubble rate \cite{Chatrabhuti:2015mws,Channuie:2019kus}, as well as logarithmic-corrected $R^2$ and Einstein-Hu-Sawicki models, with results compared to Planck 2018 data \cite{Waeming:2020rir}. Additionally, slow-roll inflation in $f(T)$ Gravity's Rainbow has also been investigated, with detailed calculations of primordial perturbations showing compatibility with Planck 2018 and BICEP/Keck 2021 results \cite{Leyva:2021fuo}.  

Modified gravity theories, such as $f(R)$ Gravity's Rainbow, serve as effective field theories that bridge classical GR and quantum gravity by incorporating low-energy quantum corrections. However, fully understanding the quantum nature of gravity requires a proper quantisation framework. The ADM formalism \cite{Arnowitt:1962hi} introduced canonical quantisation for gravity by constructing an infinite-dimensional superspace of $3D$ spatial metrics. In 1983, Hartle and Hawking \cite{Hartle:1983ai} refined this approach by restricting the superspace to finite-dimensional minisuperspaces, simplifying the problem and making possible the formulation of the Wheeler-DeWitt equation. The solution to this equation, known as the Wave Function of the Universe, provides probability amplitudes for configurations and predicts classical trajectories.  

Despite these advancements, the canonical quantisation procedure does not yield a fully renormalizable theory of gravity within the framework of Quantum Field Theory. Moreover, interpretations of the Wave Function, such as Everett's Many-Worlds \cite{Everett:1957hd} and Hawking's probabilistic approach \cite{Hartle:1983ai}, remain topics of ongoing debate. Nonetheless, quantum cosmology’s focus on simplified models has proven more practical than the full quantum gravity framework, providing tractable solutions and valuable insights into the very early universe, inflation, and singularity resolution \cite{Bojowald:2011zzb}. Canonical quantum cosmology may also lead to observable effects via quantum corrections to primordial spectra, potentially leaving testable imprints in the cosmic microwave background (CMB) \cite{Chataignier:2023rkq}.

In this work, we investigate the quantum dynamics of $f(R)$ Gravity's Rainbow in the presence of a perfect fluid, employing Schutz's formalism to establish a well-defined notion of time \cite{Schutz:1970my,Schutz:1971ac}. While previous works \cite{Vakili:2009he} have examined the quantum cosmology of $f(R)$ Gravity with a Schutz perfect fluid, we extend this analysis by incorporating rainbow effects into the spacetime metric within the $f(R)$ Gravity's Rainbow framework. Through canonical quantisation, we recover the notion of time and derive the Schr\"{o}dinger–Wheeler–DeWitt (SWD) equation for the quantum model. By solving its eigenfunctions, we construct the wave function of the Universe. We focus on stiff matter in the very early universe as the perfect fluid and obtain analytical solutions for the wave function in this specific scenario.

The paper is structured as follows: Section \ref{Rainbow_G} introduces Rainbow Geometry, while Section \ref{fRainbow_Shutz} presents $f(R)$ Gravity’s Rainbow in Schutz's formalism. Section \ref{Class_Dyncs} explores the classical dynamics, deriving the cosmological field equations and obtaining both analytical and numerical solutions. Section \ref{Cano_quant} is dedicated to the canonical quantisation of the model. Finally, in Section \ref{Conclusions} we give the conclusions. 

Throughout this work, we adopt natural units where $c=\hbar=16\pi G=1$.

\section{Rainbow geometry}\label{Rainbow_G}

Magueijo and Smolin \cite{Magueijo:2002xx} proposed extending doubly special relativity to curved spacetimes, introducing a semiclassical quantum gravity framework where the Planck energy or length acts as a universal constant alongside the speed of light. In this theory, non-linear Lorentz transformations in momentum space result in a modified dispersion relation (MDR),
\be
E^2 f_{1}^2(E)-\mathbb{P}^2 f_{2}^2(E)=\mathit{m}_{0}^2,
\label{MDR}
\ee where $f_{1}(E)$ and $f_{2}(E)$ depend on energy $E$, and $\mathbb{P}$ is the particle momentum. Null geodesics reveal that the speed of light becomes energy-dependent, $c(E) = f_{2}(E) / f_{1}(E)$. Consequently, the modified equivalence principle implies that spacetime is described by an energy-dependent rainbow metric. 

The spacetime metric for the Friedmann-Lema\^{i}tre-Robertson-Walker (FLRW) universe with rainbow effect is written as \cite{Magueijo:2002xx}
\bea
ds^2=-\frac{\mathcal{N}(t)^2}{f_{1}(E)^2} dt^2+\frac{a(t)^2}{f_{2}(E)^2}\left[\frac{dr^2}{1-K r^2}+r^2 d\Omega^2\right],
\eea where $\mathcal{N}(t)$ is the lapse function, $a(t)$ the scale factor, and $K=-1, 0, 1$ corresponds to the open, flat, and closed universe respectively. $d\Omega^2=d\theta^2+\sin{\theta}^2 d\phi^2$ is the element of solid angle. Thus, $f_{1}(E)$ and $f_{2}(E)$ are functions of the energy of the probe particle. According to the correspondence principle \cite{Magueijo:2002xx}, the rainbow functions $f_{1}(E)$ and $f_{2}(E)$ satisfy the limit conditions $f_{1}(E)\rightarrow{1}$ and $f_{2}(E)\rightarrow{1}$ for $E\rightarrow E_{Pl}$, with $E_{Pl}$ the Planck energy.

\section{$f(R)$ Gravity's Rainbow in Schutz's formalism}\label{fRainbow_Shutz}

The relevant action is given by
{\small
\bea
&& S=S_{g}+S_{m}=\int_{M}{d^{4}x\sqrt{-g} f(R)}+ 2\int_{\partial M}{d^{3}x\sqrt{h}h_{a b} K^{a b}}+\nonumber\\
&& \int_{M}{d^{4}x\sqrt{-g} p},
\label{TheAction}
\eea} where $f(R)$ is a function of the curvature scalar $R$ associated to Levi-Civita connection, $K^{a b}$ is the extrinsic curvature and $h_{a b}$ is the three-dimensional spatial induced metric. Furthermore, the last term of \eqref{TheAction} represents the
matter contribution to the total action where $p$ is the pressure of the perfect fluid which is related to its energy density by the equation of state (EoS)
\be
p=w \rho.
\ee
In Schutz’s formalism \cite{Schutz:1970my,Schutz:1971ac}, the fluid's four-velocity is expressed in terms of five potentials $\epsilon$, $\zeta$, $\beta$, $\theta$ and $S$ as
\be
U_{\nu}=\frac{1}{\mu}\left(\epsilon_{,\nu}+\zeta \beta_{,\nu}+\theta S_{,\nu}\right),
\ee where $_{,\nu}\equiv \partial /\partial x^{\nu}$, and $U_{\nu}$ satisfies the condition
\be
U_{\nu}U^{\nu}=-1,
\ee The variables $\mu$ and $S$ are the specific enthalpy and specific entropy, respectively. Moreover, the potentials $\zeta$ and $\beta$ are related to torsion and therefore are absent in FRW models, whereas $\epsilon$ and $\theta$ have no clear physical interpretation in this formalism.

In the minisuperspace, the gravitational part of the action is written as 
{\small
\bea
&& S_{g}=\int{dt \mathcal{L}_{g}},\nonumber\\
&&=\int dt\Bigg\{ \frac{\mathcal{N} a^3}{f_{1} f_{2}^{3}} f(R)- \zeta\Bigg[R-\frac{6}{\mathcal{N}^2}\Bigg(\frac{f_{1}^2\ddot{a}}{a}+\frac{f_{1}^2\dot{a}^2}{a^2}+\frac{K f_{2}^{2} \mathcal{N}^2}{a^2}-\nonumber\\
&& \frac{f_{1}^2 \dot{\mathcal{N}} \dot{a}}{\mathcal{N}a}\Bigg)\Bigg]\Bigg\}, 
\eea} where the overdot denotes differentiation  with respect to cosmic time, and $\zeta$ is a Lagrange multiplier which is obtained by variation with respect to $R$, that is, $\zeta=\mathcal{N} a^3 f_{,R}/(f_{1} f_{2}^3)$, with $f_{,R}\equiv df/dR$. 
So, one obtains the following point-like Lagrangian
for the gravitational part of the model
\bea
&& \mathcal{L}_{g}=-\frac{6 f_{,R}}{\mathcal{N}}\frac{f_{1}}{f_{2}^3} a \dot{a}^2- \frac{ 6 f_{,RR}}{\mathcal{N}} \frac{f_{1}}{f_{2}^3} \dot{R} a^2 \dot{a}+\frac{6 K \mathcal{N}  f_{,R}}{f_{1} f_{2}} a+ \nonumber\\
&& \frac{\mathcal{N} a^3}{f_{1} f_{2}^{3}}\left(f-f_{,R} R\right),
\eea where $f_{,RR}\equiv d^2f/dR^2$. Also, by defining $\phi=f_{,R}$, we obtain $\dot{\phi}=f_{,RR} \dot{R}$,  allowing us to rewrite the previous Lagrangian as
{\small
\bea
&& \mathcal{L}_{g}=-\frac{6}{\mathcal{N}} \frac{f_{1}}{f_{2}^{3}} a \dot{a}^2 \phi-\frac{6}{\mathcal{N}} \frac{f_{1}}{f_{2}^3} a^2 \dot{a}  \dot{\phi}+\frac{6 K \mathcal{N}}{f_{1} f_{2}} a \phi  -\frac{\mathcal{N} a^3}{f_{1} f_{2}^3} V(\phi),
\eea} where $V(\phi)=R f_{,R}-f=R \phi-f$.

On the other hand, in Schutz's formalism \cite{Schutz:1970my,Schutz:1971ac}, the matter Lagrangian density takes the form \cite{Majumder:2013ypa}
\be
\mathcal{L}_{m}=\frac{\mathcal{N} a^3}{f_{1} f_{2}^{3}} p=\frac{f_{1}^{\frac{1}{w}}}{f_{2}^{3}} \frac{w \mathcal{N}^{-\frac{1}{w}} a^{3}}{\left(1+w\right)^{1+\frac{1}{w}}}\left(\dot{\epsilon}+\theta \dot{S}\right)^{1+\frac{1}{w}} e^{-\frac{S}{w}}.
\ee

We are now ready to construct the Hamiltonian for the model. In terms of the conjugate
momenta, the super-Hamiltonian is given by
\bea
\mathscr{H}=\dot{a} P_{a} +\dot{\phi} P_{\phi}+\dot{\epsilon} P_{\epsilon}+
\dot{S}P_{S}-\mathcal{L},
\label{H1}
\eea where $\mathscr{H}=\mathscr{H}_{g}+\mathscr{H}_{m}$ and $\mathcal{L}=\mathcal{L}_{g}+\mathcal{L}_{m}$. The momenta conjugate corresponding to each of the variables can
be derived using the definition $P_{q}=\frac{\partial \mathcal{L}}{\partial \dot{q}}$, which yields 
\bea
\label{Pa}
&& P_{a}=\frac{\partial \mathcal{L}}{\partial \dot{a}}=-\frac{12}{\mathcal{N}}\frac{f_{1}}{f_{2}^{3}} a \dot{a} \phi -\frac{6}{\mathcal{N}}\frac{f_{1}}{f_{2}^{3}} a^2 \dot{\phi},\\
\label{Pphi}
&& P_{\phi}=\frac{\partial \mathcal{L}}{\partial \dot{\phi}}=-\frac{6}{\mathcal{N}}\frac{f_{1}}{f_{2}^{3}} a^2 \dot{a},\\
&& P_{\epsilon}=\frac{\partial \mathcal{L}}{\partial \dot{\epsilon}}=\frac{f_{1}^{\frac{1}{w}}}{f_{2}^{3}} \mathcal{N}^{-\frac{1}{w}} a^3 \frac{\left(\dot{\epsilon}+\theta \dot{S}\right)^{\frac{1}{w}}}{\left(1+w\right)^{\frac{1}{w}}} e^{-\frac{S}{w}},\\
&& P_{S}=\theta P_{\epsilon}.
\eea
Using the relations
\bea
&& T=-P_{S} e^{-S} P_{\epsilon}^{-(1+w)},\\
&& P_{T}=P_{\epsilon}^{1+w} e^{S},
\eea
and applying the canonical transformation, Eq. \eqref{H1} leads to 
\bea
&&   \mathscr{H} =\mathcal{N}\mathcal{H},\nonumber\\
&& =\mathcal{N}\Bigg[-\frac{f_{2}^{3}}{6 f_{1}}\frac{P_{a} P_{\phi}}{a^{2}}+
\frac{f_{2}^{3}}{6 f_{1}} \frac{\phi}{a^{3}} P_{\phi}^2-\frac{6 K a \phi}{f_{1} f_{2}}+\frac{a^{3} V(\phi)}{f_{1} f_{2}^{3}}+\nonumber\\
&& \frac{f_{2}^{3 w}}{f_{1}}\frac{P_{T}}{a^{3 w}}\Bigg].
\label{Hamiltonian}
\eea

The momentum $P_{T}$, as the only remaining canonical variable associated with matter, appears linearly in the Hamiltonian, completing the construction of the phase space and the formulation of the model's Lagrangian and Hamiltonian.  

The classical dynamics is governed by the Hamiltonian equations, that is
\bea
\label{dota1}
&& \dot{a}=\Big\{a,\mathscr{H}\Big\}=-\frac{\mathcal{N}}{6}\frac{f_{2}^3}{f_{1}} \frac{P_{\phi}}{a^2},\\
\label{dotPa1}
&& \dot{P}_{a}=\Big\{P_{a},\mathscr{H}\Big\}=\mathcal{N}\Bigg[-\frac{f_{2}^3}{3 f_{1}}\frac{P_{\phi} P_{a}}{a^3}+\frac{f_{2}^{3}}{2 f_{1}} \frac{\phi}{a^4} P_{\phi}^2+\nonumber\\
&& \frac{6 K \phi}{f_{1} f_{2}}-\frac{3 a^2 V(\phi)}{f_{1} f_{2}^{3}}+\frac{3 w f_{2}^{3 w}}{f_{1}} a^{-(1+3 w)} P_{T}\Bigg],\\
\label{dotphi1}
&& \dot{\phi}=\Big\{\phi,\mathscr{H}\Big\}=\mathcal{N}\left[-\frac{f_{2}^{3}}{6 f_{1}}\frac{P_{a}}{a^2}+\frac{f_{2}^{3}}{3 f_{1}}\frac{\phi}{a^3} P_{\phi}\right],\\
\label{dotPphi1}
&& \dot{P}_{\phi}=\Big\{P_{\phi},\mathscr{H}\Big\}=\mathcal{N}\left[-\frac{f_{2}^{3}}{ 6 f_{1}} \frac{P_{\phi}^2}{a^3}+\frac{6 K a}{f_{1} f_{2}}-\frac{a^3 V_{,\phi}}{f_{1} f_{2}^3}\right],\\
\label{dotT}
&& \dot{T}=\Big\{T,\mathscr{H}\Big\}=\frac{f_{2}^{3 w}}{f_{1}} \frac{\mathcal{N}}{a^{3 w}},\\
\label{dotPT}
&& \dot{P}_{T}=\Big\{P_{T},\mathscr{H}\Big\}=0,
\eea where we have defined $V_{,\phi}\equiv dV/d\phi$. The constraint equation $\mathcal{H} = 0$  reveals the under-determined nature of the cosmological model concerning the concept of time. This issue can be addressed at the classical level by fixing the gauge \cite{Bojowald:2011zzb}. By choosing the gauge $\mathcal{N} = \frac{f_{1}}{f_{2}^{3w}} a^{3w}$, the variable $T$ becomes equivalent to $t$, enabling $T$ to act as the time parameter. As a result, the classical equations of motion can be reformulated in this gauge as follows
\bea
\label{dota2}
&&\dot{a}=-\frac{1}{6} f_{2}^{3\left(1-w\right)} a^{3 w-2} P_{\phi},\\
\label{dotPa2}
&&\dot{P}_{a}=-\frac{1}{3} f_{2}^{3 \left(1-w\right)} a^{3\left(w-1\right)} P_{\phi} P_{a}+\frac{1}{2} f_{2}^{3 \left(1-w\right)} a^{3 w-4} \phi P_{\phi}^{2}+\nonumber\\
&& \frac{6 K \phi a^{3 w}}{f_{2}^{1+3 w}}-\frac{3 a^{3 w+2} V(\phi)}{f_{2}^{3 \left(1+w\right)}}+3 w a^{-1} P_{0},\\
\label{dotphi2}
&& \dot{\phi}=-\frac{1}{6} f_{2}^{3 \left(1-w\right)} a^{3 w-2} P_{a} +\frac{1}{3} f_{2}^{3 \left(1-w\right)} a^{3\left(w-1\right)} \phi P_{\phi},\\
\label{dotPphi2}
&& \dot{P}_{\phi}=-\frac{1}{6} f_{2}^{3 \left(1-w\right)} a^{3 \left(w-1\right)} P_{\phi}^{2} +\frac{6 K a^{3 w+1}}{f_{2}^{3 w+1}}-\nonumber\\
&& \frac{a^{3\left(w+1\right)} V_{,\phi}}{f_{2}^{3 \left(1+w\right)}},
\eea where we have set $P_{T} = P_{0} = \text{const}$ from  Eq. \eqref{dotPT}. Since the system's integrability depends on the choice of $f(R)$, which determines the potential $V(\phi)$, we first focus on this aspect. 

\section{Classical dynamics}\label{Class_Dyncs}
For $\mathcal{N}=(f_{1}/f_{2}^{3w})a^{3w}$,  and using equations \eqref{Pa} and \eqref{Pphi}, one obtains 
\bea
&& P_{a}=-12 f_{2}^{3\left(w-1\right)}a^{1-3w}\dot{a} \phi-6 f_{2}^{3\left(w-1\right)}a^{2-3w} \dot{\phi},\\
&& P_{\phi}=-6 f_{2}^{3\left(w-1\right)}a^{2-3w} \dot{a}.
\eea 

In this case, the background equations \eqref{dotPa2} and \eqref{dotPphi2} take the form 
{\small
\bea
\label{dotPa3}
&& \ddot{\phi}+(2-3 w)H \dot{\phi}+ \left[2 \dot{H}+3(1-2w)H^2+K\left(\frac{a}{f_{2}}\right)^{2\left(3 w-1\right)}\right]\phi-\nonumber\\
&& \frac{1}{2} \left(\frac{a}{f_{2}}\right)^{6 w} V+\frac{w P_{0}}{2} \left(\frac{a}{f_{2}}\right)^{3\left(w-1\right)} =0,\\
&& \frac{\ddot{a}}{a}+ \left(1-3 w\right)H^2-\frac{1}{6} \left(\frac{a}{f_{2}}\right)^{6 w}V_{,\phi}+K  \left(\frac{a}{f_{2}}\right)^{2\left(3 w-1\right)}=0,
\label{dotPphi3}
\eea}
respectively, where $H\equiv \dot{a}/a$ is the Hubble rate. The other background equations correspond to identities. 

On the other hand, the Hamiltonian constraint $\mathcal{H}=0$ leads us to
{\small
\bea
&& P_{0}=6\left(\frac{a}{f_{2}}\right)^{3 (1-w)} H \dot{\phi}+6 \left(\frac{a}{f_{2}}\right)^{3 (1-w)} H^2 \phi-\left(\frac{a}{f_{2}}\right)^{3 (1+w)} V+\nonumber\\
&& 6 K \phi \left(\frac{a}{f_{2}}\right)^{1+3 w}.
\label{H_zero}
\eea}
Combining  equations \eqref{dotPa3} and \eqref{dotPphi3}, and using \eqref{H_zero}, we obtain
{\small
\bea
&&\ddot{\phi}-H \left[4 H \phi+(3 w+1) \dot{\phi}\right]+\frac{1}{3} \left(\frac{a}{f_{2}}\right)^{6w}  \phi V_{,\phi}+\nonumber\\
&&\frac{1}{2} (w+1) \left(\frac{a}{f_{2}}\right)^{3(w-1)}  P_{0}-4 K  \phi  \left(\frac{a}{f_{2}}\right)^{2\left(3 w-1\right)}=0.
\label{dotPphi_stiff2}
\eea }Note that Eq. \eqref{H_zero} corresponds to the first Friedmann equation for the model in Schutz's formalism. The parameter $P_{0}$ plays the role of a cosmological constant.  Additionally, Eq. \eqref{dotPphi3} corresponds to the second Friedmann equation, while Eq. \eqref{dotPphi_stiff2} represents the equation of motion for the effective field $\phi$. However, only two of these three equations are independent. 

By introducing the number of $e$-folds $N\equiv \log(a)$, Eq. \eqref{H_zero} transforms into
\be
H^2=\frac{P_{0}}{6 \left(\phi'+\phi\right)}\left(\frac{e^{N}}{f_{2}}\right)^{3(w-1)}+ \frac{V-6 \left(\frac{f_{2}}{e^{N}}\right)^2 K \phi}{6 \left(\phi'+\phi\right)}\left(\frac{e^{N}}{f_{2}}\right)^{6w},
\label{Fr00}
\ee while Eq. \eqref{dotPphi3} is rewritten as 
\be
H'=(3 w-2) H+ \frac{\left(\frac{e^N}{f_{2}}\right)^{6 w} \left[V_{,\phi}-6 K \left(\frac{f_{2}}{e^{N}}\right)^2\right]}{6 H},
\ee where the prime denotes differentiation with respect to $N$. Moreover, the equation of motion for the field \eqref{dotPphi_stiff2} becomes
{\smaller
\bea
&&\phi''-4 \phi+\mathcal{B}\Bigg\{3\left[ P_{0} w+\left( \phi V_{,\phi}-V-4 K \phi \left(\frac{f_{2}}{e^{N}}\right)^2  \right)\left(\frac{e^N}{f_{2}}\right)^{3\left( w+1\right)} \right]\phi'+\nonumber\\
&& 3 P_{0} (w+1) \phi + \left(V_{,\phi}-6 \left(\frac{f_{2}}{e^{N}}\right)^2 K\right)\phi'^2 \left(\frac{e^N}{f_{2}}\right)^{3\left( w+1\right)} +\nonumber\\
&& 2 \left(V_{,\phi}-12 K \left(\frac{f_{2}}{e^{N}}\right)^2 \right) \phi^2  \left(\frac{e^N}{f_{2}}\right)^{3\left( w+1\right)}\Bigg\}=0,
\label{dotPphi_stiff3}
\eea} where $\mathcal{B}=\frac{1}{P_{0}+ \left( V-6 K \phi \left(\frac{f_{2}}{e^{N}}\right)^2 \right)\left(\frac{e^N}{f_{2}}\right)^{3\left( w+1\right)}}$.

Let us consider the function \cite{Starobinsky,DeFelice:2010aj}
\be
f(R)=\eta R+\lambda R^{\upsilon},
\label{f(R)_ansatz}
\ee where $\eta$, $\lambda$ and $\upsilon$ are constants. Thus, we have $\phi=f_{,R}=\eta +\lambda \upsilon R^{\upsilon-1}$ and then $R=\left(\frac{\phi-\eta}{\lambda \upsilon}\right)^{1/(\upsilon-1)}$. Therefore, the scalar potential can be written as 
\be
V(\phi)=\lambda (\upsilon-1) \left(\frac{\phi-\eta}{\lambda \upsilon}\right)^{\upsilon/\left(\upsilon-1\right)}.
\label{S_Pot}
\ee 
 For Starobinsky inflation \cite{Starobinsky}, with $\upsilon = 2$, the effective potential takes the form 
\be
V(\phi) = \frac{\left(\phi - \eta\right)^{2}}{4\lambda}.
\label{Pot}
\ee
By introducing the variable $\mathcal{U}(\phi)=\phi'$, and setting $P_{0}=K=0$, the motion equation for the field simplifies to
\be
\mathcal{U}_{,\phi}=3 -\frac{6 \phi }{\phi-\eta }-\frac{4 \phi^2}{(\phi-\eta ) \mathcal{U}}-\frac{2 \mathcal{U}}{\phi-\eta }+\frac{4 \phi }{\mathcal{U}}. 
\label{MEqphiSimp}
\ee 
In order to obtain some analytical results, we perform the transformation $\phi=\eta+\varphi$. Thus, we consider the two limits: $\varphi \gg \eta$ and $\varphi \ll \eta$.

In the first case ($\varphi \gg \eta$), Eq. \eqref{MEqphiSimp} is reduced to
\be
\mathcal{U}_{,\varphi}=-3-\frac{2 \mathcal{U}}{\varphi},
\ee whose solution is 
\be
\mathcal{U}(\varphi)=\varphi'= \frac{C_1}{\varphi^2}-\varphi,
\ee where $C_{1}$ is an integration constant. The real-valued solution to the above equation is given by
\be
\phi\sim \varphi(N)= \left[C_{1}+e^{-3 (C_{2}+N)}\right]^{\frac{1}{3}}.
\ee Since the condition $\varphi\gg \eta$ holds, one has $N\ll-\frac{1}{3} \log \left(\eta ^3-C_{1}\right)-C_{2}$.

Therefore, from Eq. \eqref{Fr00}, one obtains
\be
H(N)=\frac{\left[C_{1}+e^{-3 (C_{2}+N)}\right]^{\frac{2}{3}}}{2  \sqrt{6 C_{1}\lambda }}\left(\frac{e^{N}}{f_{2}(E)}\right)^{3w}.
\ee This allows us to calculate
\be
\epsilon_H(N)\equiv -\frac{H'}{H}=-3w +\frac{2}{1+C_{1} e^{3( C_{2}+ N)}}.
\ee The parameter $\epsilon_{H}$ is related to the deceleration parameter through $\epsilon_{H}=1+q$. Thus, accelerated expansion occurs when $\epsilon_{H}<1$, or equivalently, when $q<0$. For $0<C_{1}<\eta^3 $ and $1/3\leq  w\leq 1$, accelerated expansion always takes place. 

On the other hand, in the limit $\varphi\ll \eta$, one has
\be
\mathcal{U}_{,\varphi}=-\frac{2 \mathcal{U}}{\varphi}-\frac{6 \eta }{\varphi}-\frac{4 \eta ^2}{\varphi \mathcal{U}}.
\ee
Additionally, assuming $\mathcal{U}\gg\eta$, the solution to the above equation is written as
\be
\mathcal{U}= \frac{\tilde{C}_{1}}{3\varphi^2}, 
\ee and thus
\be
\varphi (N)=\left(\tilde{C}_{2}+\tilde{C}_{1} N\right)^{\frac{1}{3}},
\ee where $\tilde{C}_{1}$ and $\tilde{C}_{2}$ are integration constants. So, the condition $\varphi\ll \eta\ll \mathcal{U} $, with $\eta>0$, implies $N\ll\frac{\eta ^3}{\tilde{C}_{1}}-\frac{\tilde{C}_{2}}{\tilde{C}_{1}}$ for $\tilde{C}_{1}>3\eta^3$, and $N \ll -\frac{\tilde{C}_{2}}{\tilde{C}_{1}} +\frac{1}{3}\sqrt{\frac{\tilde{C}_{1}}{3 \eta^3}}$ for $0<\tilde{C}_{1}<3 \eta^3$.

In this case, the Hubble parameter becomes
\be
H(N)=\frac{(\tilde{C}_{1} N+\tilde{C}_{2})^{\frac{2}{3}}}{2 \sqrt{2 \tilde{C}_{1}\lambda }}\left(\frac{e^{N}}{f_{2}(E)}\right)^{3 w},
\ee and thus
\be
\epsilon_{H}(N)=-\frac{2 \tilde{C}_{1}}{3 (\tilde{C}_{1} N+\tilde{C}_{2})}-3 w. 
\ee Accelerated expansion occurs for $-\frac{\tilde{C_{2}}}{\tilde{C_{1}}}-\frac{2}{3(3 w+1)}<N<-\frac{\tilde{C_{2}}}{\tilde{C_{1}}}$, where $\eta>0$, $\tilde{C}_{1}>0$ and $0\leq w\leq 1$.

\begin{figure*}[ht]
\centering 
\includegraphics[width=0.48\textwidth]{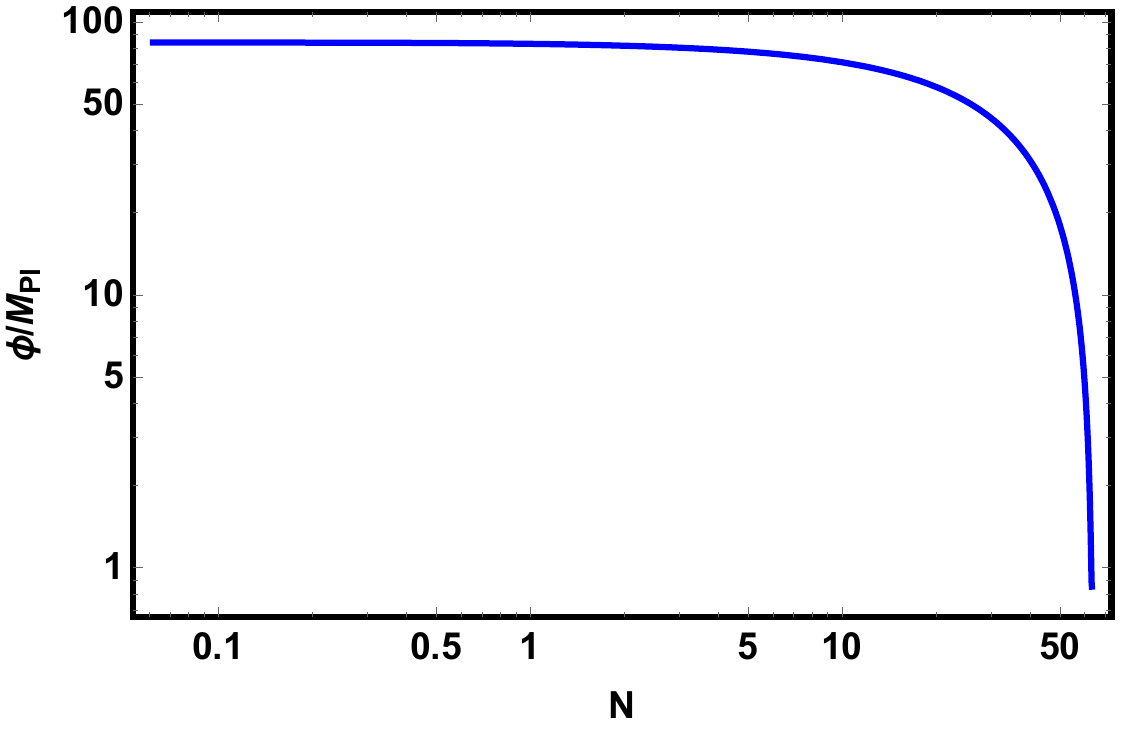}
\includegraphics[width=0.5\textwidth]{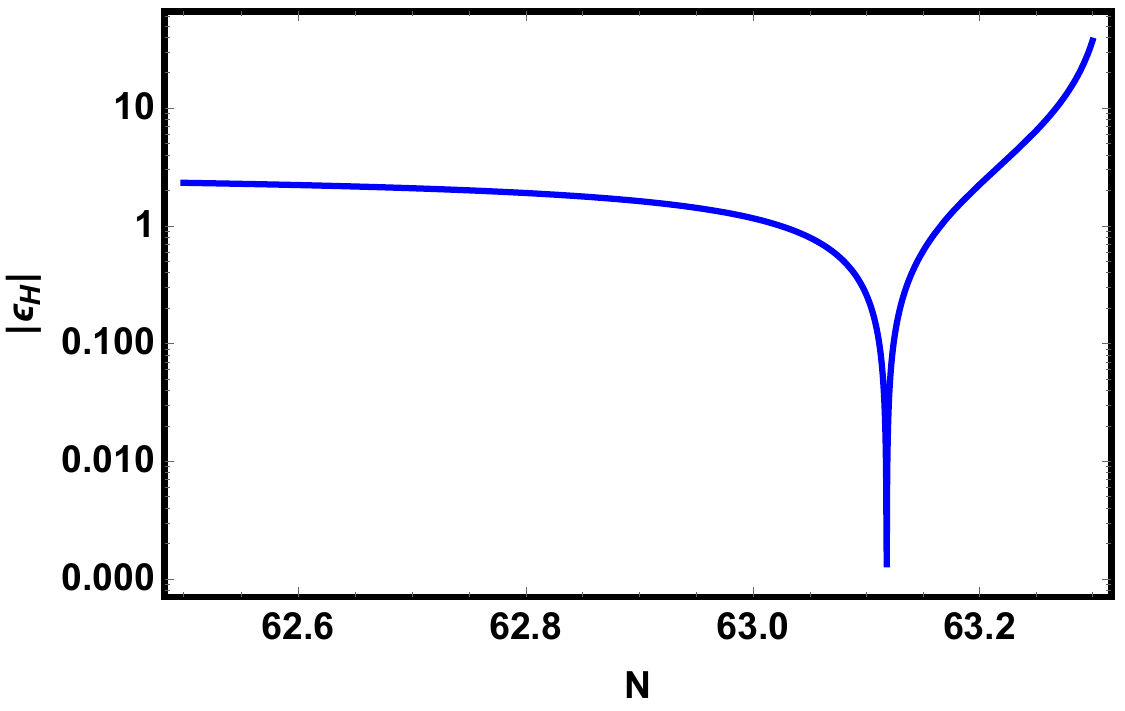} \\ 
\caption{The left panel shows the evolution of the scalar field $\phi(N)$, while the right panel depicts 
$|\epsilon_{H}|=|H'/H|$, both for $w=1$ and $\eta=1$ and the initial conditions $\phi(0)/M_{Pl}=84$ and $\mathcal{U}(0)\equiv \phi'(0)/{M_{Pl}}=0.05$.
}
\label{fig1_Classical}
\end{figure*}

The analytical results can be corroborated via numerical integration. By numerically solving Eq.~\eqref{MEqphiSimp} along with $\phi' = \mathcal{U}$, we obtain $\phi(N)$ and subsequently $\epsilon_{H}(N)$. Figure~\ref{fig1_Classical} shows their evolution for $\eta = 1$, $w = 1$, and the same initial conditions. This choice of EoS is natural in the context of the very early universe.  Before inflation begins, a massive inflaton field undergoes an ultra-hard EoS ($w= p/\rho \approx 1$) \cite{MukhanovBook}, consistent with the expectation that, in a high-density Universe, the sound speed approaches the speed of light $c_{s}=\sqrt{\partial p/\partial \rho}\approx 1$
 \cite{Xu:2016rdf}. The left panel shows the behaviour of the scalar field $\phi(N)$, which starts at a large value and gradually decreases. The right panel depicts the evolution of $\epsilon_{H}(N)$, which remains negative until a few e-folds before the end of inflation, indicating a phase of super-inflation where $H' > 0$ (or equivalently, $\dot{H} > 0$). 
 
 Interestingly, the occurrence of super-inflation is not unique to this model; a similar effect also appears in Loop Quantum Cosmology due to quantum gravity corrections near the Planck scale~\cite{Ashtekar:2006wn}. It is worth noting that although the dynamics of the scalar field in terms of the number of e-folds $N$ becomes independent of rainbow effects in the absence of a cosmological constant and spatial curvature, the Hubble rate  $H$ remains influenced by both the potential and rainbow modifications. Specifically, the rainbow function $f_2(E)$ appears explicitly in Eq.~\eqref{Fr00}, with its functional form determined by the underlying rainbow gravity model \cite{Magueijo:2002xx}.

\section{Canonical Quantisation}\label{Cano_quant}
We study the quantum cosmology of the model by deriving the Wheeler–DeWitt equation from the Hamiltonian \eqref{Hamiltonian}. With the lapse function $\mathcal{N}$ acting as a Lagrange multiplier, the Hamiltonian constraint $\mathcal{H} = 0$ requires, under Dirac quantisation, that the quantum states of the universe be annihilated by the operator form of $\mathcal{H}$:
\bea
&& \mathcal{H} \Psi=\Bigg[-\frac{f_{2}^{3}}{6 f_{1}}\frac{P_{a} P_{\phi}}{a^{2}}+
\frac{f_{2}^{3}}{6 f_{1}} \frac{\phi P_{\phi}^2}{a^{3}} -\frac{6 K a \phi}{f_{1} f_{2}}+ \frac{a^{3} V(\phi)}{f_{1} f_{2}^{3}}+\nonumber\\
&& \frac{f_{2}^{3 w}}{f_{1}}\frac{P_{T}}{a^{3 w}}\Bigg] \Psi=0. 
\label{WDWEq}
\eea The wave function of the universe, $\Psi(a, \phi, T)$, introduces a factor-ordering issue when constructing the corresponding quantum mechanical operator equation. Special care must be taken when replacing the dynamical variables $q$ and $P_q$ with their quantum operator counterparts to ensure Hermiticity. To address this, the operator form of equation \eqref{WDWEq} is constructed with proper ordering considerations \cite{Vakili:2009he}
{\small
\bea
&&\Bigg[-\frac{1}{12}\left(a^{n} P_{a} a^{m}+a^{m} P_{a} a^{n}\right) P_{\phi}+ \frac{1}{12 a^3}\Big(\phi^{s} P_{\phi} \phi^{u} P_{\phi} \phi^{v}+\nonumber\\
&& \phi^{v} P_{\phi} \phi^{u} P_{\phi} \phi^{s}\Big)-\frac{6 K a \phi}{f_{2}^4}+ \frac{a^{3} V(\phi)}{f_{2}^{6}}+\frac{f_{2}^{3 (w-1)}}{a^{3 w}} P_{T}\Bigg] \Psi(a,\phi,T)=0. 
\label{WDWEq2}
\eea} 
The parameters $n$, $m$, $s$, $u$, and $v$ satisfy $n + m = -2$ and $s + u + v = 1$, representing the factor-ordering ambiguity for $a$ and $P_{a}$, and $\phi$ and $P_{\phi}$ in \eqref{WDWEq2}. Substituting $P_{a} = -i\frac{\partial}{\partial a}$ and similarly for $P_{\phi}$ and $P_{T}$, the equation becomes
{\small
\bea
&& \Bigg[\frac{a^{-2}}{6}\frac{\partial^2}{\partial a\partial \phi}-\frac{a^{-3}}{3}\frac{\partial}{\partial \phi}-\frac{a^{-3}}{6} \phi \frac{\partial^2}{\partial \phi^2}+\frac{A a^{-3}}{6} \phi^{-1}- \frac{6 K a \phi}{f_{2}^{4}}+\nonumber\\
&& \frac{a^{3} V(\phi)}{f_{2}^{6}}- i f_{2}^{3 \left(w-1\right)} a^{-3 w} \frac{\partial}{\partial T}\Bigg] \Psi(a,\phi,T)=0,
\eea} where we have defined $A=sv$.
The equation adopts the form of a Schrödinger equation, $i\frac{\partial}{\partial T}\Psi = \hat{\mathcal{H}}\Psi$, with the Hamiltonian operator $\hat{\mathcal{H}}$ being Hermitian under the standard inner product 
\be
\avg{\Phi|\Psi}=\int_{(a,\phi)}{a^{-3w}\Phi^{*}\Psi da d\phi},
\ee regardless of the choice of ordering parameters \cite{Vakili:2009he}.

We separate the variables in the SWD equation
\be
\Psi(a,\phi,T)=e^{-i E T}\psi(a,\phi),
\ee where $E$ is a separation constant that, due to the assumption of a constant average energy for the probe particles, can be interpreted as the scale at which the minisuperspace is probed. This separation of variables leads to
{\small
\bea
&& \Bigg\{\frac{x^2}{4}\frac{\partial^2}{\partial x^2}-\frac{x}{4} \frac{\partial}{\partial x}-y^2 \frac{\partial^2}{\partial y^2}-2 y \frac{\partial}{\partial y}+\Bigg[A-\frac{36 K x^4}{f_{2}^4}+ \frac{6 x^6 y^{-2} V(y)}{f_{2}^6}-\nonumber\\
&& \frac{6 E x^{3 \left(1-w\right)} y^{\frac{1}{2}\left(3 w-1\right)} }{f_{2}^{3 \left(1-w\right)}}\Bigg]\Bigg\} \psi(x,y)=0, 
\label{SWDEq2}
\eea}
where we defined the variables
\be
x(a,\phi) = a \phi^{1/2}, \:\:\:\: y(a, \phi)=\phi.
\ee

To proceed further, we assume the ansatz \eqref{f(R)_ansatz}, with $\upsilon =2$, corresponding to Starobinsky's model \cite{Starobinsky}.  At very early times, it can be assumed that $\eta R \ll \lambda R^{2}$ \cite{Starobinsky,DeFelice:2010aj}, allowing us to set $\eta = 0$. Consequently, the scalar potential in Eq. \eqref{Pot} simplifies to $V(y) = \frac{y^{2}}{4\lambda}$, and Eq. \eqref{SWDEq2} takes the form
{\small
\bea
&& \Bigg\{\frac{x^2}{4}\frac{\partial^2}{\partial x^2}-\frac{x}{4} \frac{\partial}{\partial x}-y^2 \frac{\partial^2}{\partial y^2}-2 y \frac{\partial}{\partial y}+\Bigg[A-\frac{36 K x^4}{f_{2}^4}+ \frac{3 x^6}{2 \lambda f_{2}^6}-\nonumber\\
&& \frac{6 E x^{3 \left(1-w\right)} y^{\frac{1}{2}\left(3 w-1\right)} }{f_{2}^{3 \left(1-w\right)}}\Bigg]\Bigg\} \psi(x,y)=0. 
\label{SWDEq3}
\eea}
Under this assumption, equation \eqref{SWDEq3} separates into two independent equations for the specific cases of $w = 1$ (stiff matter) and $w = 1/3$ (radiation). In the following, we focus on solving for the wavefunction $\psi(x, y)$ in the stiff matter case \cite{Xu:2016rdf}.

In this case, we obtain 
{\smaller
\bea
&& \Bigg\{\frac{x^2}{4}\frac{\partial^2}{\partial x^2}-\frac{x}{4} \frac{\partial}{\partial x}-y^2 \frac{\partial^2}{\partial y^2}-2 y \frac{\partial}{\partial y}+\Bigg[A-\frac{36 K x^4}{f_{2}^4}+ \frac{3 x^6}{2 \lambda f_{2}^6}-\nonumber\\
&& 6 E y\Bigg]\Bigg\} \psi(x,y)=0.
\label{SWDEq3_Stiff}
\eea} Thus, taking $\psi(x,y)=X(x) Y(y)$ we get
{\small
\bea
\label{X}
&& \Bigg[x^2\frac{d^2}{dx^2}-x\frac{d}{dx}+\left(1-\nu^2-\frac{144 K x^4}{f_{2}^4}+\frac{6 x^6}{\lambda f_{2}^6}\right)\Bigg] X(x) =0,\\
&& \Bigg[y^2\frac{d}{dy^2}+2 y\frac{d}{dy}+6E y-\frac{\nu^2-1}{4}\Bigg]Y(y)=0. 
\label{Y}
\eea} 
Following \cite{Vakili:2009he}, we set $(\nu^2-1)/4$ as the separation constant and chose the factor-ordering parameter $A = 0$, as it does not affect semiclassical calculations in quantum cosmology \cite{Hawking:1985bk}. While the wavefunction's behaviour generally depends on $A$ \cite{Steigl:2005fk}, the substitution $\nu \rightarrow \sqrt{\nu^2 - 4A}$ in \eqref{X} or \eqref{Y} links $A$ to the separation constant. Since the total wavefunction is a superposition over all $\nu$, its key features remain independent of $A$ \cite{Vakili:2009he}. Moreover, for $A = 0$, one has $s = 0$ or $v = 0$, or both, ensuring the Hermiticity of the second term in \eqref{WDWEq2}.  
 
For a flat space, $K=0$, we obtain the following solutions
{\small
\bea
&& X(x)=x\Bigg[c_1 J_{\frac{\nu}{3}}\left(\frac{\sqrt{\frac{2}{3 \lambda}} x^3}{f_{2}(E)^3}\right)+c_2 Y_{\frac{\nu}{3}}\left(\frac{\sqrt{\frac{2}{3\lambda}} x^3}{f_{2}(E)^3}\right)\Bigg].
\eea}
and 
{\small
\bea
&& Y(y)=\frac{1}{\sqrt{y}} \Bigg[d_1 J_\nu\left(2 \sqrt{6 y E}\right)+d_2 Y_{\nu}\left(2 \sqrt{6 y E}\right)
\Bigg], 
\eea} where $c_{i}$, $d_{i}$ are constants. $J_{\alpha}(x)$ and $Y_{\alpha}(x)$ are Bessel functions of the first and second kind, respectively \cite{Arfken2013Mathematical}. Thus, the eigenfunctions are given by
\be
\Psi_{\nu E}(x,y,T)=e^{-i E T} \frac{x}{\sqrt{y}} J_{\frac{\nu}{3}}\left(\frac{\sqrt{\frac{2}{3 \lambda}} x^3}{f_{2}(E)^3}\right) J_{\nu}\left(2 \sqrt{6 y E}\right).
\ee Since well-defined functions in all ranges of variables $x$ and $y$ are desired, we have taken $c_{2}=d_{2}=0$ \cite{Vakili:2009he}. 

Therefore, the general solution to the SWD equation is obtained as the superposition of the eigenfunctions, which yields
\be
\Psi(x,y,T)=\int_{E=0}^{\infty}{\int_{\nu=0}^{\infty}{A(E)C(\nu)}\Psi_{\nu E}(x,y,T)dEd\nu}.
\ee In order to construct the wave packet, we have introduced the suitable weight functions $A(E)$ and $C(\nu)$. Since we are interested in studying the rainbow effects we assume the following linear relation
\be
f_{2}(E)=1+\kappa E, 
\ee where $0<\kappa E< 1$ and $\kappa$ is a constant. 

Thus, choosing the quasi-Gaussian weight factor \cite{Vakili:2009he}
\be
A(E)=12(24 E)^{\frac{\nu}{2}} e^{-24 \sigma E},
\ee with $\sigma$ an arbitrary positive constant, and using the relation \cite{watson1995bessel}
{\smaller
\bea
&& \int_{0}^{\infty}J_{\nu}(at)e^{-p^2 t^2}t^{\mu-1}dt=\frac{\Gamma(\frac{1}{2}\nu + \frac{1}{2}\mu) (\frac{a}{2 p})^{\nu}e^{-\frac{a^2}{4p^2}}}{2p^\mu \Gamma(\nu+1)}\times \nonumber\\
&& {}_{1}{F}_1\left(\frac{1}{2}\nu - \frac{1}{2}\mu + 1; \nu + 1; \frac{a^2}{4p^2}\right),
\eea } with $\Gamma(z)$ the Gamma function and ${}_{1}{F}_1(a;b;z)$ the confluent hypergeometric function, also known as Kummer's function,
we obtain 
\be
\Psi=\Psi_{1}+\Psi_{2},
\ee where 
{\small
\bea
&& \Psi_{1}=\frac{x e^{-\frac{y}{4\sigma+i \frac{T}{6}}}}{\sqrt{y}}\int_{0}^{\infty}\Bigg[C(\nu)J_{\frac{\nu}{3}}\left(X\right) \frac{y^{\frac{\nu}{2}}}{\left(2 \sigma+i\frac{T}{12}\right)^{\nu+1}}\Bigg] d\nu,
\eea}
and 
{\smaller
\bea
&& \Psi_{2}=\frac{xe^{-\frac{y}{4\sigma+i \frac{T}{6}}}}{\sqrt{y}}\int_{0}^{\infty}\displaystyle{\Bigg[\frac{\sqrt{\frac{2}{3 \lambda}} \kappa C(\nu) J'_{\frac{\nu}{3}} x^3 y^{\frac{\nu}{2}}\left(y-\left(4 \sigma + i \frac{T}{6}\right) (\nu+1) \right)}{\left(2 \sigma+i \frac{T}{12}\right)^{\nu+3}}\Bigg]}d\nu.
\nonumber\\
\eea} For this result, we have applied a Taylor expansion to the rainbow function and $J'_{\frac{\nu}{3}}=dJ_{\frac{\nu}{3}}\left(X\right)/dX$, with $X\equiv \sqrt{\frac{2}{3 \lambda}}x^3$. 

By performing the variable change $\nu\rightarrow 3 \nu$ and $x\rightarrow X=(\sqrt{2/3\lambda}) x^3$, we get
\be
\Psi_{1}=\frac{3}{2} \left(\frac{3\lambda}{2}\right)^{\frac{1}{6}}\frac{X^{\frac{1}{3}}}{y^{\frac{1}{2}}} e^{-\frac{y}{4 \sigma+i\frac{T}{6}}}\sum_{n=1}^{2}\int_{0}^{\infty}{\frac{C(\nu) y^{\frac{3\nu}{2}}\mathcal{H}_{\nu}^{(n)}(X)}{\left(2\sigma+i\frac{T}{12}\right)^{3\nu+1}}} d\nu,
\label{psi1_stiff}
\ee where $\mathcal{H}_{\nu}^{(n)}(X)$, $n=1,2$ are the Hankel functions of first and second kind, respectively. We have used the relation $J_{\nu}(X)=\left(\mathcal{H}_{\nu}^{(1)}(X)+\mathcal{H}_{\nu}^{(2)}(X)\right)/2$ \cite{Arfken2013Mathematical}. Similarly, we obtain
{\small
\bea
&& \Psi_{2}=\frac{3}{4}\left(\frac{3 \lambda}{2}\right)^{\frac{1}{6}}\frac{X^{\frac{4}{3}}}{y^{\frac{1}{2}}}e^{-\frac{y}{4 \sigma+i\frac{T}{6}}}\sum_{m=0}^{1}\sum_{n=1}^{2}\int_{0}^{\infty}\Bigg\{\frac{\kappa C(\nu) y^{\frac{3 \nu}{2}}}{\left(2\sigma+i\frac{T}{12}\right)^{3\left(\nu+1\right)}}\times\nonumber\\
&& (-1)^{m}\mathcal{H}^{(n)}_{\nu-1+2m}(X) \Bigg[y-\left(4\sigma+i\frac{T}{6}\right)\left(3 \nu+1\right)\Bigg]\Bigg\}d\nu,
\label{psi2_stiff}
\eea} where we have also used $J_{\nu}'(X)=\left(J_{\nu-1}(X)-J_{\nu+1}(X)\right)/2$ \cite{Arfken2013Mathematical}.

In principle, one could attempt to find a weight function 
$C(\nu)$ so that the total wave packet $\Psi$ can be analytically calculated. However, the expressions in \eqref{psi1_stiff} and \eqref{psi2_stiff} are too complicated, rendering this task a highly challenging objective. Therefore, following Ref. \cite{Vakili:2009he}, we assume a shifted Gaussian weight function $C(\nu)\sim e^{-b (\nu-s)^2}$. This is a natural choice in the context of quantum mechanics and may allow us to numerically obtain a localized wave packet, opening the possibility to extract physical results. For an arbitrary function that changes slowly over the range where the Gaussian function has significant value, and if the Gaussian is narrow enough, one can approximate 
{\smaller
\bea
&& \int_{0}^{\infty}{e^{-b (\nu-s)^2} F(\nu)d\nu}\simeq \frac{1}{2}\sqrt{\frac{\pi}{b}}\left[1+\erf({\sqrt{b}s})\right]F(s)+\nonumber\\
&& \sum_{m=1}^{\infty}\frac{C(m) b^{-\frac{(1+m)}{2}}}{2 m!}\frac{\partial^{m} F(s)}{\partial \nu^{m}},
\eea} where $C(m)=\Gamma\left(\frac{m+1}{2}\right) + (-1)^m \gamma\left(\frac{m+1}{2}, b s^2\right)$, and $\gamma$ denotes the lower incomplete gamma function \cite{Arfken2013Mathematical}.

By applying this approximation to Eqs. \eqref{psi1_stiff} and \eqref{psi2_stiff}, one obtains, at leading order
{\small
\bea
&& \Psi(X,y,T)\simeq \frac{y^{\frac{\left(3 s-1\right)}{2}}e^{-\frac{y}{4\sigma+i\frac{T}{6}}}}{\left(2\sigma +i\frac{T}{12}\right)^{3s+1}}\Bigg[C_{1}X^{\frac{1}{3}} \sum_{n=1}^{2}{\mathcal{H}^{(n)}_{s}(X)}+\nonumber\\
&& C_{2}\frac{X^{\frac{4}{3}}\left[y-\left(4\sigma+i\frac{T}{6}\right)\left(3s+1\right)\right]}{\left(2\sigma+i\frac{T}{12}\right)^{2}}\sum_{m=0}^{1}\sum_{n=1}^{2}(-1)^m\mathcal{H}^{(n)}_{s-1+2m}(X)\Bigg],
\label{FullPsi}
\eea} where $C_{1}=\frac{3}{4}\left(\frac{3\lambda}{2}\right)^{\frac{1}{6}} \sqrt{\frac{\pi}{b}}\left(1+\erf({\sqrt{b}s})\right)$ and $C_{2}=\frac{3}{8}\left(\frac{3\lambda}{2}\right)^{\frac{1}{6}} \sqrt{\frac{\pi}{b}}\kappa \left(1+\erf({\sqrt{b}s})\right)$ are constants.
We are particularly interested in analysing the system's behaviour in two limiting cases of $X$: one for small values and the other for large values.  

\begin{figure*}[ht]
\centering
\begin{tabular}{cccc}
\scriptsize \textbf{$T=0$} & \scriptsize \textbf{$T=10$} & \scriptsize \textbf{$T=30$} & \scriptsize \textbf{$T=50$} \\ 
\includegraphics[width=0.22\textwidth]{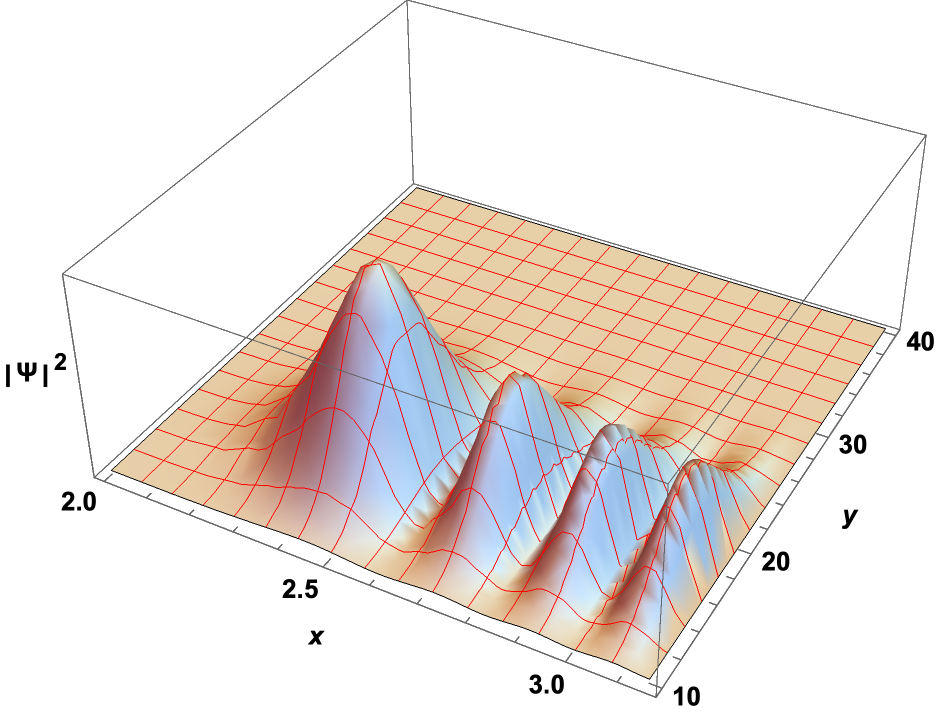} & 
\includegraphics[width=0.22\textwidth]{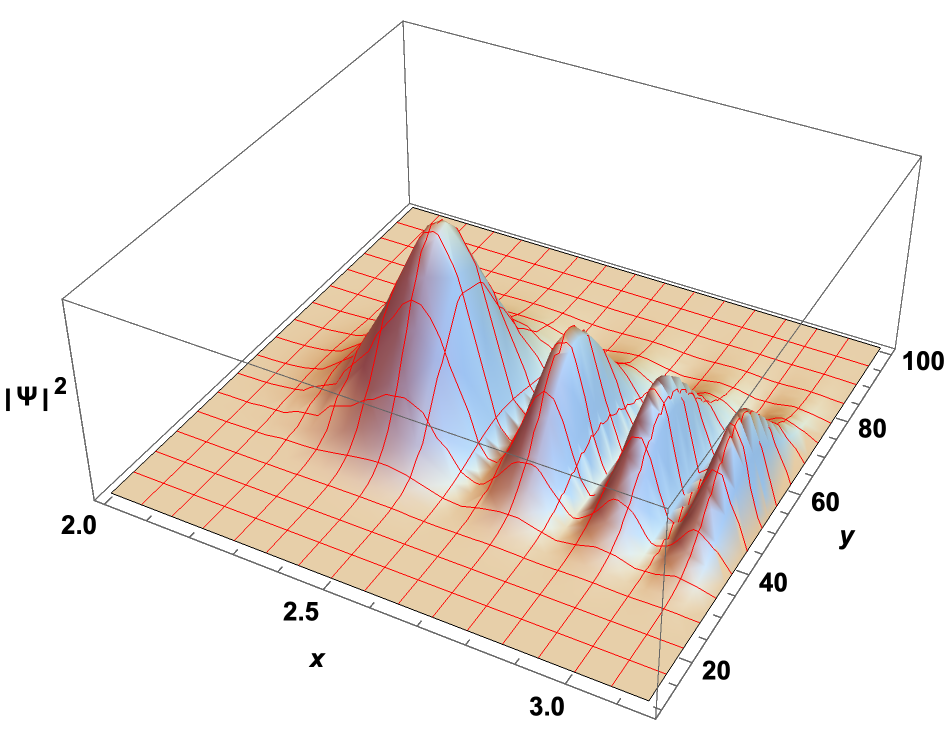} & 
\includegraphics[width=0.22\textwidth]{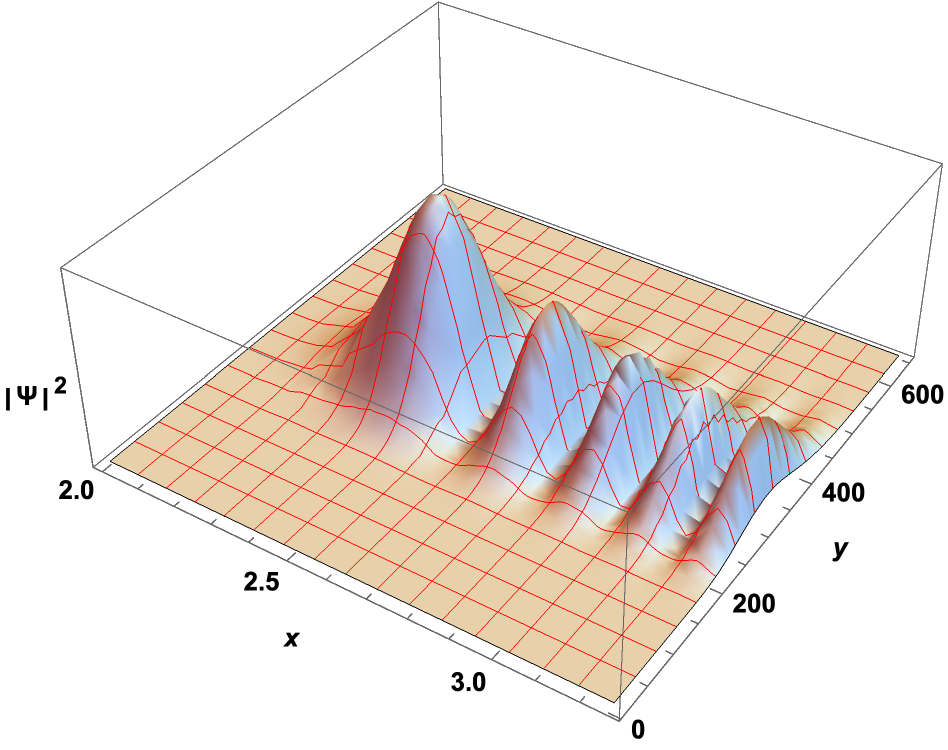} & 
\includegraphics[width=0.22\textwidth]{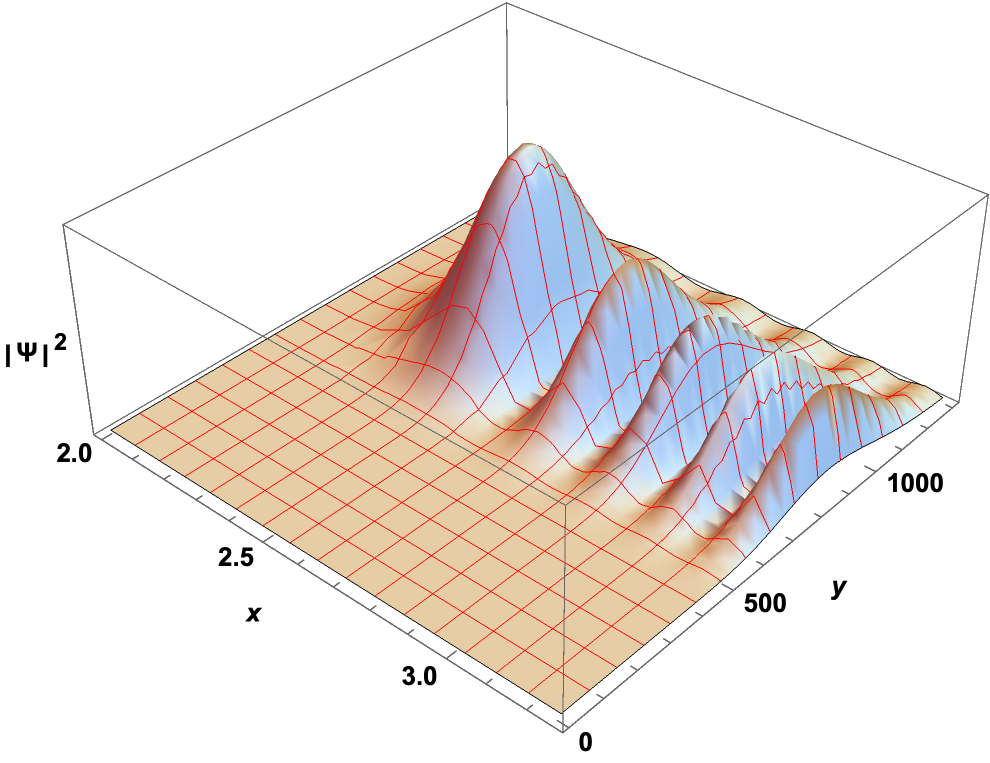} \\ 
\includegraphics[width=0.22\textwidth]{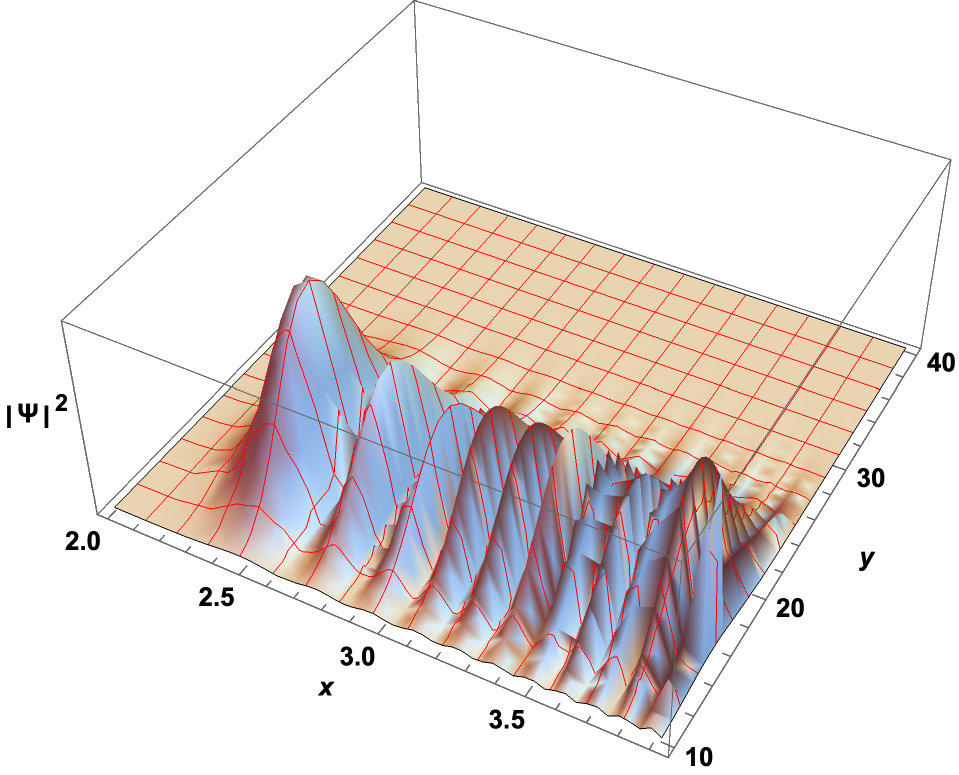} & 
\includegraphics[width=0.22\textwidth]{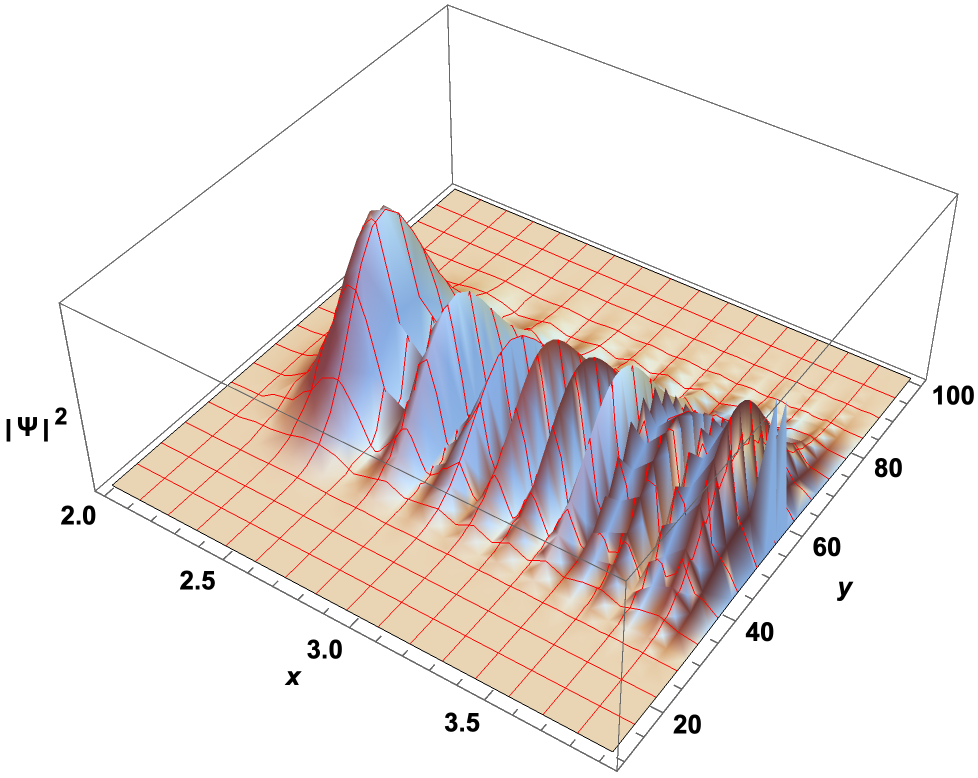} & 
\includegraphics[width=0.22\textwidth]{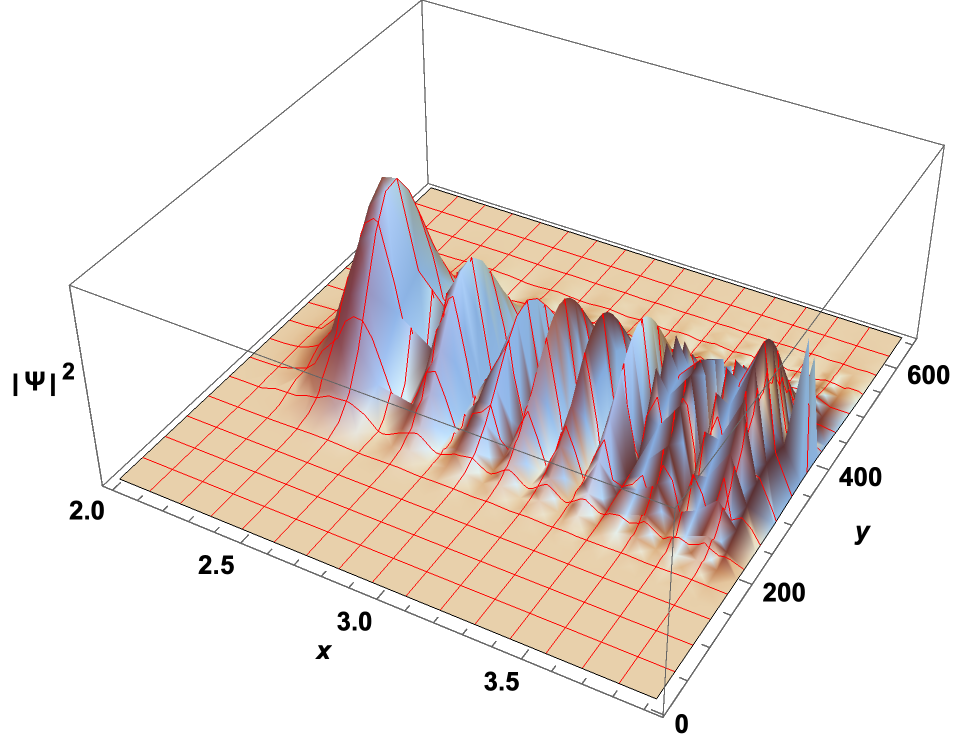} & 
\includegraphics[width=0.22\textwidth]{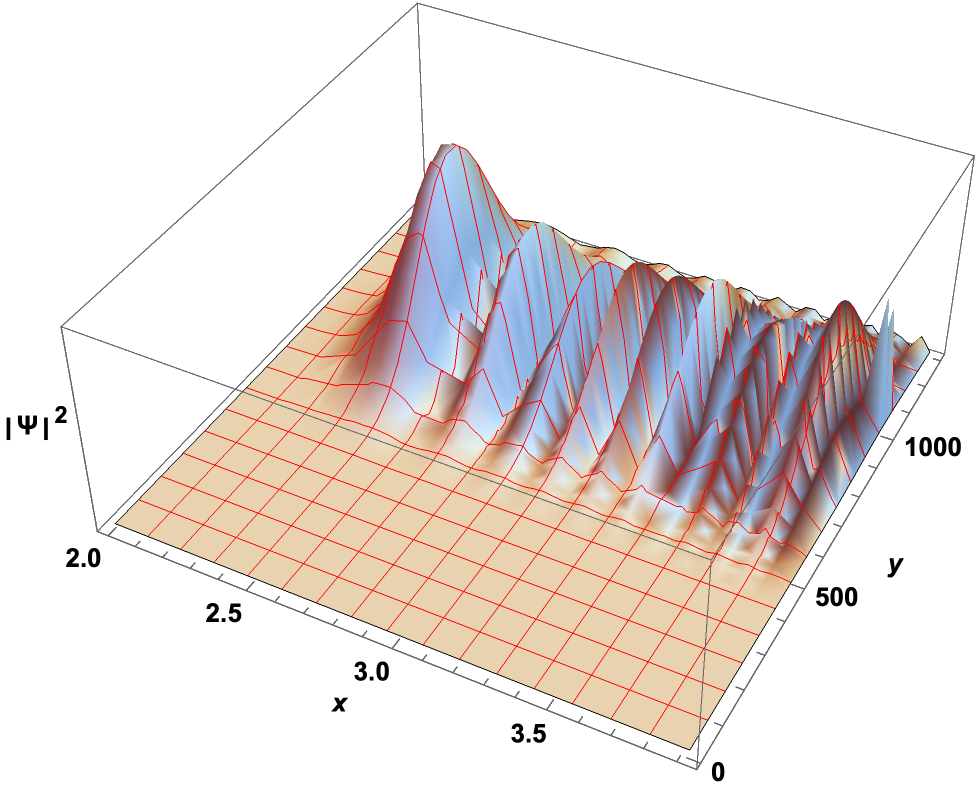} \\ 
\end{tabular}
\caption{Probability amplitude distribution at 
$T=0,10,30,50$: upper row without rainbow effects, $\kappa=0$; lower row with rainbow effects, $\kappa=5\times 10^{-4}$. Also, we have used the values $\lambda=1$, $s=10$, $\sigma=0.3$ and $b=10^3$.
}
\label{fig1}
\end{figure*}

For $X\rightarrow \infty$ one has \cite{Arfken2013Mathematical,Gonzalez-Espinoza:2020azh}
\be
H^{(n)}_{\nu}{(X)}\sim \sqrt{\frac{2}{\pi X}} e^{(-1)^{n+1}i X}e^{(-1)^{n}i \frac{\pi}{2}\left(\nu+\frac{1}{2}\right)}, 
\ee and then 
\bea
&& \Psi(X,y,T)\simeq \frac{y^{\frac{\left(3s-1\right)}{2}}e^{-\frac{y}{4\sigma+i\frac{T}{6}}}}{\left(2\sigma +i\frac{T}{12}\right)^{3s+1}}\Bigg[C_{1}\frac{\cos{\left[X-\frac{\pi}{2}\left(s+\frac{1}{2}\right)\right]}}{X^{\frac{1}{6}}}+\nonumber\\
&& C_{2}\frac{X^{\frac{5}{6}}\left[\left(4\sigma+i\frac{T}{6}\right)\left(3s+1\right)-y\right]\sin{\left[X-\frac{\pi}{2}\left(s+\frac{1}{2}\right)\right]}}{\left(2\sigma+i\frac{T}{12}\right)^{2}}\Bigg].
\label{BigX}
\eea 
where $C_2=\kappa C_{1}/2$.\\
On the other hand, for $X\rightarrow 0$ one has \cite{Arfken2013Mathematical,Gonzalez-Espinoza:2020azh}
\be
H^{(n)}_{\nu}{(X)}\sim \left(-1\right)^{n}i\frac{\Gamma{\left(\nu\right)}}{\pi} \left(\frac{2}{X}\right)^{\nu}, 
\ee and then 
\be
\Psi(X,y,T)\approx 0.
\label{SmallX}
\ee
In FIGS. \ref{fig1}, we use Eq. \eqref{FullPsi} to depict the behaviour of the wave packet within the ${x,y}$ space. These plots corroborate the analytical results described by Eq. \eqref{BigX} and \eqref{SmallX}, which correspond to the limiting cases $X \to \infty$ and $X \to 0$, respectively. The upper row shows the probability amplitude distribution at different times ($T=0, 10, 30, 50$) without rainbow effects, while the lower row presents the same scenario with rainbow effects ($\kappa=5\times 10^{-4}$). Each peak in the plot represents a probable quantum state of the universe, characterized by its geometry (scale factor) and the effective scalar field $y=\phi=f_{,R}$, which encapsulates the modifications to gravity at a given time. 

At $T=0$, the wavefunction's dominant peak near specific nonzero values of $x$ and $y$ suggests the universe likely emerged from this state, with smaller peaks indicating alternative quantum states that could interact via tunneling, representing possible transitions between multiple universes in the past \cite{Vakili:2009he}. The evolution of these peaks over the internal time $T$, defined via the perfect fluid in Schutz's formalism, describes the dynamic interaction between the universe's geometry and modified gravitational effects in the quantum regime. The collective behaviour of these peaks over time reveals the probabilistic evolution of the universe within a quantum cosmological framework.

As time progresses, the wave packet propagates predominantly in the $y$-direction, its width increasing and its peaks shifting with a group velocity toward larger 
$y$ values. Given that $y=\phi$, this evolution suggests that the universe transitions to states with increasing values of $R$ over time \cite{Vakili:2009he}. Notably, even in the presence of rainbow effects, this directional propagation---already observed in their absence---remains unchanged.

In the absence of rainbow effects, the largest peak consistently appears at small $X$, regardless of time $T$, indicating that the quantum cosmological model favors configurations with a small scale factor and a growing effective scalar field. In contrast, when rainbow effects are present, the peaks grow at larger $X$, resulting in a distinct behaviour at large scales. This suggests that the rainbow effects favor configurations with a larger scale factor while still preserving the increasing behaviour of $\phi$ over time. 

In order to obtain the predictions for the universe from this quantum model, we adopt the many-worlds interpretation \cite{Alvarenga:2001nm,Vakili:2009he}. The expectation value of the scale factor is calculated as 
\bea
&& \avg{a}(T)=\frac{\int{a^{-3w}\Psi(X,y,T)^{*}a\Psi(X,y,T)}dad\phi}{\int{a^{-3w}\Psi(X,y,T)^{*}\Psi(X,y,T) dad\phi}}=\nonumber\\
&& \alpha\frac{\int{X^{-\frac{4}{3}} y^\frac{1}{2}|\Psi|^2 dXdy}}{\int{X^{-\frac{5}{3}} y |\Psi|^2 dXdy}},
\eea 

with $\alpha=\left(\frac{3\lambda}{2}\right)^{1/6}$, and 
\bea
&& \avg{\phi}(T)=\frac{\int{a^{-3w}\Psi(X,y,T)^{*}\phi\Psi(X,y,T)}dad\phi}{\int{a^{-3w}\Psi(X,y,T)^{*}\Psi(X,y,T) dad\phi}}=\nonumber\\
&& \frac{\int{X^{-\frac{5}{3}}y^2 |\Psi|^2 dXdy}}{\int{X^{-\frac{5}{3}}y |\Psi|^2 dXdy}},
\eea where we have assumed stiff matter $w=1$. 

\begin{figure*}[ht]
\centering
\includegraphics[width=0.375\textwidth]{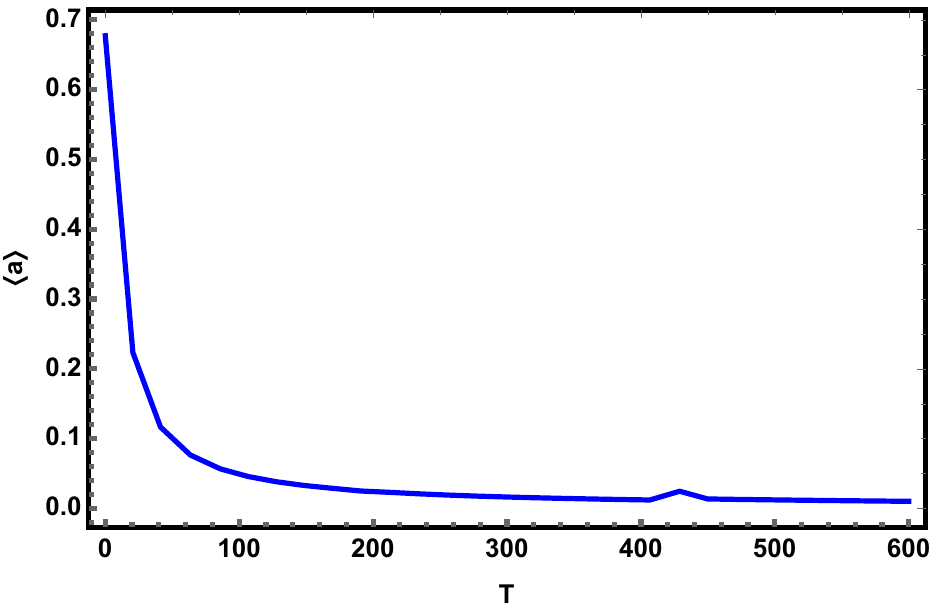} 
\includegraphics[width=0.40\textwidth]{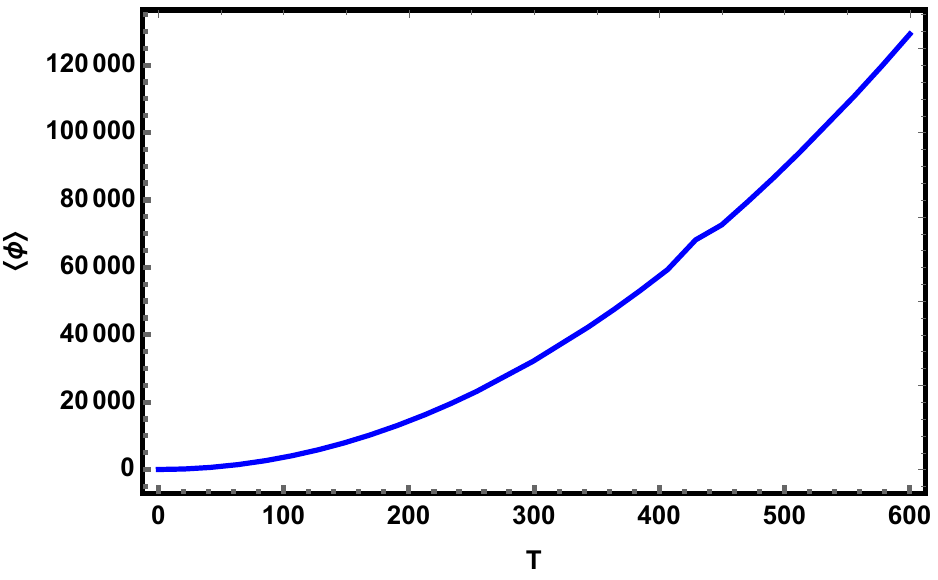} 
\includegraphics[width=0.40\textwidth]{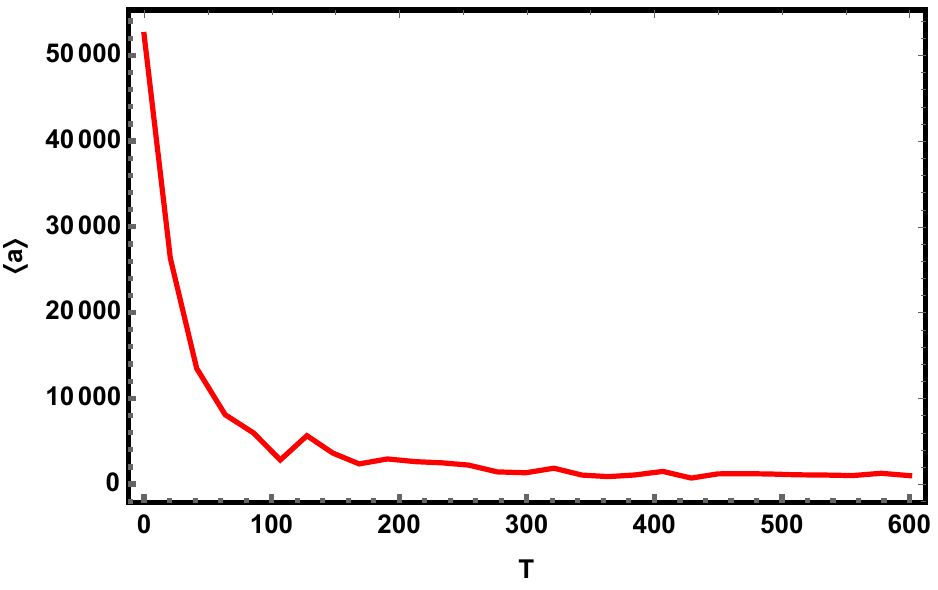}  
\includegraphics[width=0.40\textwidth]{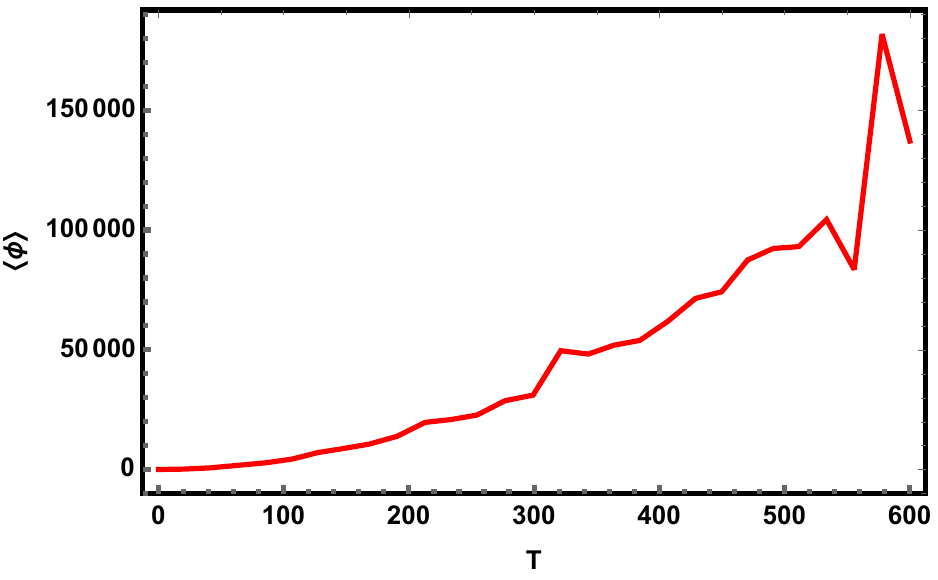} \\ 
\caption{Evolution of the expected values of the scale factor and the effective scalar field for parameter values $\lambda=1$, $s=10$, $\sigma=0.3$ and $b=10^3$. The upper row (blue lines) represents the evolution  of $\avg{a}(T)$  and  $\avg{\phi}(T)$ in the absence of rainbow effects ($\kappa=0$), while the lower row (red lines) shows their evolution in the presence of rainbow effects with $\kappa=5\times 10^{-4}$.
}
\label{fig2}
\end{figure*}

By using \eqref{FullPsi}, in FIG. \ref{fig2}, we depict the evolution of $\avg{a}$(T) and $\avg{\phi}$(T) in the absence (upper row) and presence (lower row) of the rainbow effect, respectively.  In both scenarios, the quantum nature of the model prevents the occurrence of the initial singularity, resulting in a contracting phase that appears to lead to a bounce  \cite{Brandenberger:2002ty,Peter:2002cn}. However, the presence of rainbow corrections strengthens this transition, as indicated by the significantly larger initial values of $\avg{a}$. Additionally, at later times, the expected value of the effective scalar field $\avg{\phi}$ grows more rapidly in the presence of rainbow corrections, suggesting that these quantum gravity modifications may influence the transition from the quantum to the classical regime.

In the classical limit,  the expected value $\avg{\phi}$ can be identified with the inflaton field, implying that rainbow gravity effects may modify the initial conditions for inflation, particularly the initial field value $\phi_{i}=\phi(t_{i})$ at the onset of inflation. Such modifications could impact the total $e$-folding number $N$, the energy scale of inflation, and the inflationary observables. In large-field inflation models (e.g., chaotic and monodromy inflation), a higher $\phi_{i}$ typically leads to a larger field excursion $\Delta{\phi}\equiv \left|\phi_{e}-\phi_{i}\right|$ which, according to the Lyth bound $\Delta{\phi}/M_{Pl}\gtrsim \mathcal{O}(1)\times \left(r/0.01\right)^{1/2}$ \cite{Lyth:1998xn,Brandenberger:2016uzh}, implies a higher tensor-to-scalar ratio $r$ and enhances the prospects of detecting primordial gravitational waves in future CMB experiments. However, note that the exact form of the Lyth bound depends on the underlying gravity theory, meaning that the relationship between $\Delta{\phi}$ and $r$ may be modified in alternative frameworks \cite{Odintsov:2023aaw,Yang:2015pga,Gao:2014pca}. Moreover, in small-field models (e.g., Starobinsky and Higgs inflation), the effects are more model-dependent, as a larger $\phi_{i}$ does not necessarily lead to a higher $N$ or a larger inflationary energy scale. In particular, for flat potentials, even if $N$ increases, $r$ may remain small \cite{Ellis:2023wic}. 

Recent studies \cite{Leyva:2022zhz,Channuie:2024sba,Corda:2010uq,Barrow:2013gia,Amelino-Camelia:2013wha} indicate that rainbow gravity can alter the parameter $r$, keeping it within observational bounds ($r<0.036$ \cite{BICEP:2021xfz}) while still allowing for potential detection in future CMB experiments. These modifications may alter the Lyth bound, impacting the relationship between the inflaton field excursion $\Delta{\phi}$ and  $r$. Upcoming observations from BICEP/Keck \cite{BICEP:2021xfz} and LiteBIRD \cite{LiteBIRD:2022cnt,LiteBIRD:2024wix} will be crucial in testing these effects and assessing the role of rainbow gravity in inflationary dynamics.

\section{Concluding remarks}\label{Conclusions}

In this work, we investigated the classical and quantum cosmology of  $f(R)$  gravity’s rainbow in the presence of a perfect fluid, employing Schutz’s formalism to recover a well-defined notion of time. We derived the cosmological field equations governing the classical dynamics of the model and obtained analytical solutions in certain asymptotic limits. These analytical results were further corroborated by numerical integrations for Starobinsky's model, assuming a perfect fluid behaving as stiff matter. Additionally, through canonical quantisation, we formulated the Schr\"{o}dinger–Wheeler–DeWitt (SWD) equation and obtained analytical solutions for the wave function of the Universe, calculating the expected values of the fields. 

In the classical regime, we identified the emergence of a super inflationary phase, a phenomenon also observed in other scenarios due to quantum gravity corrections \cite{Ashtekar:2006wn}. Notably, while the scalar field dynamics in terms of the number of $e$-folds remains unaffected by rainbow effects in the absence of a cosmological constant and spatial curvature, the Hubble rate continues to be influenced by both the potential and modifications introduced by rainbow gravity. This dependence arises explicitly through the rainbow function $f_{2}(E)$, which appears in the modified Friedmann equations. 

Within the framework of quantum cosmology, we found that the quantum nature of the model can remove the classical singularity, leading to a contracting phase that transitions into a bounce. This is evident from the evolution of the expectation values $ \avg{a}$  and  $\avg{\phi}$, which indicate that the Universe smoothly evolves through a minimum size without encountering a singularity. Moreover, the presence of the rainbow effect modifies the quantum evolution by altering the peak structure of the wave function, thereby affecting the probability distribution of different cosmological configurations. Specifically, we found that while the quantum dynamics still supports a bouncing scenario, the inclusion of rainbow functions increases the probability amplitude at larger scales, suggesting deviations from classical-like behaviour in the semiclassical limit. In this sense, $\avg{\phi}$ also exhibits a faster growth at later times in the presence of rainbow corrections, implying that quantum gravity effects may influence the quantum-to-classical transition.

The application of the many-worlds interpretation allowed us to extract meaningful predictions from the wave function. The quantum probability distribution suggests a dominant preference for configurations with a small scale factor and strong modified gravity corrections in the quantum regime. However, when rainbow gravity effects are included, the modified behaviour at larger scales indicates that these quantum modifications remain relevant even at later stages of evolution, potentially impacting the transition to classical cosmology.

Our analysis is based on the ADM formalism applied to a spatially homogeneous and isotropic FLRW background, resulting in a minisuperspace model suitable for quantum cosmology. We would like to highlight that our Hamiltonian framework is a symmetry–reduced version of the general ADM formulation of $f(R)$ gravity, as developed in Ref.~\cite{Capozziello:2011et}. That work constructs the complete Hamiltonian structure for $f(R)$ theories via a $(3+1)$ decomposition, identifying the constraint structure and degrees of freedom in a general setting. In our case, this formalism is adapted to a reduced phase space with a perfect fluid introduced via Schutz's method, allowing for the recovery of a time variable and enabling a Schr\"{o}dinger-like quantisation.

Additionally, the incorporation of Gravity's Rainbow deformations through energy-dependent rainbow functions modifies the effective minisuperspace geometry, leading to a deformed kinetic structure in the Hamiltonian and Wheeler–DeWitt equation. These rainbow functions act as energy-scale-dependent modifications of the canonical variables and influence both the classical dynamics and quantum evolution of the model. Despite the symmetry reduction, our model remains consistent with the general ADM framework for higher-order gravity theories.

A formal similarity exists between our model and the quantum treatment of $f(R)$ gravity presented in Ref.~\cite{Capozziello:2007gm}, where the cosmological constant emerges as an eigenvalue of the Wheeler–DeWitt equation through a one-loop quantisation approach applied to a Schwarzschild background. In that framework, the eigenvalue corresponds to vacuum energy resulting from quantum fluctuations of the gravitational field. In contrast, our model employs a minisuperspace FLRW background and introduces a perfect fluid via Schutz's formalism to define an internal time parameter. This leads to a conserved momentum $P_T = P_0$, which appears in both the classical and quantum Hamiltonians. At the classical level, this conserved quantity behaves effectively as a cosmological constant in the modified Friedmann equations. In the quantum theory, the same $P_T$ enters the Wheeler–DeWitt equation as an eigenvalue associated with the time evolution of the wave function. While this eigenvalue structure formally resembles that in Ref.~\cite{Capozziello:2007gm}, its physical interpretation is distinct: in our case, the eigenvalue arises from the matter sector rather than quantum fluctuations of spacetime geometry. Establishing a more direct correspondence would require incorporating one–loop quantum gravitational corrections, as well as a specific and consistent treatment of vacuum energy, which both lie beyond the scope of the present work. We consider this an interesting direction for future investigation.

An additional direction worth exploring involves the role of Noether symmetries in our extended minisuperspace model. In vacuum $f(R)$ cosmology, such symmetries have been used to constrain the form of the gravitational Lagrangian, reduce the dynamical system, and select specific solutions at the quantum level~\cite{Capozziello:1999xr}. The inclusion of matter through the Sorkin-Schutz formalism, however, enlarges the configuration space and modifies the symmetry conditions. This may alter the set of admissible symmetries, especially in the presence of additional fluid degrees of freedom. While this could restrict certain geometric symmetries, it may also allow new conserved quantities to emerge from cyclic variables in the matter sector. Whether such Noether symmetries persist–or can be generalized–in the presence of Rainbow functions and Schutz-type fluids remains an open question. This presents a promising line of inquiry for further investigations into the integrability and quantum structure of $f(R)$ gravity's rainbow.

Overall, our findings reinforce the idea that quantum gravity corrections and modified gravity effects play a crucial role in early-universe dynamics. The resolution of the singularity through quantum effects, further enhanced by rainbow gravity, influences the cosmological bounce and may impact the evolution of the inflaton field and inflationary observables. Furthermore, the persistence of rainbow effects at later stages suggests potential implications for the quantum-to-classical transition. Future research should further explore these quantum modifications, their impact on cosmic inflation, and their potential observational signatures in the CMB and large-scale structure.

\section*{Acknowledgments}
A.B.V. acknowledges  financial support from the University of Tarapacá. M.G-E. acknowledges the financial support of FONDECYT de Postdoctorado, N° 3230801. G.O. gratefully acknowledges the hospitality of the {\it Institute of Cosmology and Gravitation (ICG)} at the University of Portsmouth, where part of this work was carried out. G.O. would also like to thank Tiberiu Harko for valuable discussions and suggestions.




\bibliographystyle{spphys}       

\bibliography{bio}

\end{document}